\documentclass[12pt,twoside]{article}

\usepackage{paralist}
\usepackage{enumerate} %% to check paralist doesn't clash

\usepackage{amsmath,amssymb,amsthm,amscd,graphicx,color,accents}

\usepackage[T1,hyphens]{url}
\usepackage[colorlinks=true,linkcolor=blue,urlcolor=blue]{hyperref}

\usepackage{caption,lipsum,breqn}

\numberwithin{equation}{section}

\theoremstyle{plain}
\newtheorem{theorem}{Theorem}
\newtheorem*{proposition*}{Proposition}
\newtheorem{proposition}[theorem]{Proposition}
\newtheorem*{example*}{Example}

\def\be{{\boldsymbol{e}}}

\def\bLambda{\boldsymbol{\Lambda}}
\def\bnabla{\boldsymbol{\nabla}}
\def\bsigma{\boldsymbol{\sigma}}
\def\bSigma{\boldsymbol{\Sigma}}
\def\bs{\boldsymbol{s}}

\def\Cov{{{\hskip1pt}\rm{Cov}}}
\def\Var{{{\hskip1pt}\rm{Var}}}

\begin{document}

\title{\bf\large{
Statistical Implications of the Revenue Transfer Methodology in the Affordable Care Act
}}

\author{
{Michelle Li}\thanks{Schreyer Honors College, Pennsylvania State University, University Park, PA 16802.} 
\ {\ and Donald Richards}\thanks{Department of Statistics, Pennsylvania State University, University Park, PA 16802.
%\endgraf
%\ $^\ddag$Corresponding author; e-mail address: richards@stat.psu.edu
}}

\bigskip

%\today

\maketitle

\begin{abstract} 
The Affordable Care Act (ACA) includes a permanent revenue transfer methodology which provides financial incentives to health insurance plans that have higher than average actuarial risk.  In this paper, we derive some statistical implications of the revenue transfer methodology in the ACA.  We treat as random variables the revenue transfers between individual insurance plans in a given marketplace, where each plan's revenue transfer amount is measured as a percentage of the plan's total premium.  We analyze the means and variances of those random variables, and deduce from the zero sum nature of the revenue transfers that there is no limit to the magnitude of revenue transfer payments relative to plans' total premiums.  Using data provided by the American Academy of Actuaries and by the Centers for Medicare \& Medicaid Services, we obtain an explanation for empirical phenomena that revenue transfers were more variable and can be substantially greater for insurance plans with smaller market shares.  We show that it is often the case that an insurer which has decreasing market share will also  have increased volatility in its revenue transfers.  

\medskip 
\noindent 
{{\em Key words and phrases}: Affordable Care Act; Obamacare; Revenue transfers.}

\smallskip 
\noindent 
{{\em 2010 Mathematics Subject Classification}. Primary: 62P05; Secondary: 60E05.
\smallskip
\noindent
{\em JEL Classification System}.  I13, G22.}
%
%\smallskip
%\noindent
%{\em Running head}: Risk Adjustment Payments in the Affordable Care Act.
\end{abstract}

%\medskip

\section{\large Introduction}
\label{sec:introduction}

The Patient Protection and Affordable Care Act, known colloquially as the ``Affordable Care Act'' (ACA) or ``Obamacare,'' provided several measures to assist insurers with the financial expenses of meeting the new health insurance requirements.  One such measure, included in the ACA permanently, is a revenue transfer methodology to promote the equitable acceptance of risk by all insurance plans in a given marketplace.  These revenue transfers were to be done by means of a zero sum procedure which reallocated funds from health insurance plans that have experienced lower than average actuarial risk to plans that have experienced higher than average actuarial risk (Kautter et al. 2014, p.~1).  

It was later observed empirically, however, that the magnitude of revenue transfers, where each transfer payment is measured as a percentage of a plan's total premium, often were greater for insurance plans having smaller market share than for plans with larger market share.  It was also noticed that the volatility of revenue transfers often was greater for smaller plans than for larger plans (American Academy of Actuaries 2016,~p.~10; Centers for Medicare \& Medicaid Services 2016, p.~97).  These phenomena led to financial strain for some small insurance plans in Connecticut, Illinois, Maryland, Massachusetts, New Mexico, Oregon, and other states; and we refer to Herman (2016a, 2016b), Gantz (2016), Bryant (2017), Teichert (2016) and others for accounts of the ensuing financial and legal struggles, and subsequent insolvency, of some plans.  

In this paper, we provide theoretical explanations for these empirical phenomena.  We deduce, purely from the zero-sum nature of the revenue transfer methodology, that there is no limit to the magnitude of revenue transfers relative to a plan's premiums.  By treating revenue transfers as random variables and investigating their means, we deduce that mean revenue transfers can be far greater, relative to premiums, for smaller plans than for larger plans.  Second, by studying the variances of revenue transfers, we explain why the volatility of those amounts can be greater for smaller plans than for larger plans.  

Third, we apply elementary linear algebra and calculus to determine conditions under which the variance of a given plan's revenue transfer is certain to increase as the plan's market share decreases, and we show how to construct examples for which decreasing market share leads to increasing variability in revenue transfers.  For smaller plans, the generally high-magnitude revenue transfers and the increased volatility of those payments will place them at increased risk of insolvency, as was the outcome for some small plans in several of the aforementioned states.  

Our results apply, beyond the Affordable Care Act, to any health insurance system in which risk equalization is based on zero-sum revenue transfer procedures.

\section{\large The revenue transfer system and some of its implications}
\label{sec:thertf}

Let $n$ denote the number of insurance plans in a given ACA marketplace.  For $i=1,\ldots,n$, denote by $s_i$ the market share of the $i$th plan.  Clearly, $s_i > 0$ for all $i$, and $s_1+\cdots+s_n = 1$ because all plans in a marketplace jointly enroll that marketplace's entire ACA membership.  

For $i=1,\ldots,n$, we denote by $T_i$ the {\it revenue transfer} per member month for the $i$th plan, the transfer amounts being calculated as a percentage of total premiums.  Numerous actuarial factors are used to define $T_i$ (see, e.g., Patient Protection and Affordable Care Act 2013, pp.~15430--15432; Pope et al. 2014, pp.~E3-E5 and E19).  The payment amounts $T_1,\ldots,T_n$ are defined so as to reward equitable risk acceptance by all insurance plans.  A positive value of $T_i$ corresponds to a payment to the $i$th insurance plan when that plan has experienced higher than average actuarial risk, and negative values of $T_i$ correspond to a charge imposed on the $i$th plan when that plan has experienced lower than average actuarial risk.

The revenue transfer procedure is designed to be a ``zero sum'' system. Specifically, the weighted sum of all transfer payments equals zero: 
\begin{equation}
\label{zerosum}
\sum_{i=1}^n s_i T_i = 0.
\end{equation}
This zero sum requirement ensures that plans receiving payments are balanced by charges to other plans, so the total amount of funds flowing between insurance plans is unchanged (Pope et al. 2014).  

Although the revenue transfer methodology and the zero-sum condition is intended to enable equitable actuarial risk across all insurance plans in a marketplace, it is the individual values of the $T_i$ which are likely to be of primary interest to specific insurance plans.  Therefore, we discuss the set of possible values of the $T_i$ which arise from Eq.~(\ref{zerosum}).  

Consider a simple case in which there are three plans in an insurance marketplace.  If $s_1$ and $T_1$ are specified then we obtain from Eq.~(\ref{zerosum}) the equation, 
$$
s_2 T_2 + s_3 T_3 = -s_1 T_1,
$$
where $s_2, s_3 > 0$, and $s_2 + s_3 = 1 - s_1$.  For any value of $s_2$ such that $0 < s_2 < 1-s_1$ and for any value of $T_2$, we can determine from the latter equation a value for $T_3$.  

Therefore, if there are three or more plans in a marketplace then no uniform bound on the magnitude of revenue transfers for all other plans can be determined from a knowledge of the revenue transfer payment for a single plan.  This lack of a uniform bound subjects individual plans to potentially unlimited revenue transfers.  Consequently, lesser capitalized insurance plans are likely to be at greater risk of insolvency.  

We assume that the market shares $s_1,\ldots,s_n$ are deterministic.  We also 
%; equivalently, our analysis is conditional on specified values of 
%insurance plans' market shares.  
assume that each $T_i$ is a random variable and that it has finite mean and variance.  
Denote by $\mu_i$ the expected value of $T_i$, $i=1,\ldots,n$; taking expectations in Eq.~(\ref{zerosum}) and solving for $\mu_n$, we obtain 
\begin{equation}
\label{mun}
\mu_n = -\frac{1}{s_n} \sum_{i=1}^{n-1} s_i \mu_i.
\end{equation}
Suppose that $s_n$ is small.  For given values of $\mu_1,\ldots,\mu_{n-1}$ and given values of $s_1,\ldots,s_{n-1}$, the outcome of division in Eq.~(\ref{mun}) by $s_n$ may result in a number, $\mu_n$, whose magnitude is large.  This shows that an insurer with small market share will be at risk of having a highly negative revenue transfer amount, in which case the insurer will be at great risk of insolvency.  (We also note that a limit on the magnitude of revenue transfers was requested by a group, Consumers for Health Options, Insurance Coverage In Exchanges in States (2016, p.~14) ``to avoid financial harm to insurers''.)  

These observations are consistent with empirical data.  In a publication by the American Academy of Actuaries (2016, p.~10), Figure 4 in that report is a scatterplot of revenue transfers {\it vs.} insurance plan market shares for 2014, which shows that plans with small market shares often had revenue transfer amounts of very large magnitudes.   A discussion paper from the Centers for Medicare \& Medicaid Services (2016, p.~97, Figures 5.2 and 5.3) shows that smaller insurers can have highly negative revenue transfer payments, particularly so in the Small Group Market.

\section{\large Implications for the variance of revenue transfers}
\label{sec:implicationsofsigmann}

We denote $\Var(T_i)$, the variance of $T_i$, by $\sigma_{i,i}$, $1 \le i \le n$.  Also, we denote $\Cov(T_i,T_j)$, the covariance between $T_i$ and $T_j$, by $\sigma_{i,j}$, $1 \le i < j \le n$.  To analyze the variance of revenue transfers, we obtain the following result.

\begin{proposition}
\label{varTn}
The variance of $T_n$, the revenue transfer for the $n$th insurance plan, is given by
\begin{equation}
\label{sigmann}
\sigma_{n,n} = \frac{1}{s_n^2} \bigg(\sum_{i=1}^{n-1} s_i^2 \sigma_{i,i} + 2 \sum_{1 \le i < j \le n-1} s_i s_j \sigma_{i,j}\bigg).
\end{equation}
\end{proposition}

\noindent{\it Proof.}  
By Eq.~(\ref{zerosum}), 
\begin{equation}
\label{TnandotherTs}
-s_n T_n = \sum_{j=1}^{n-1} s_j T_j.
\end{equation}
We now calculate the variance of each side of this equation, applying to the right-hand side a well-known formula for the variance of a linear combination of random variables (Ross 2006, p.~357).  Then, we obtain 
\begin{align*}
s_n^2 \sigma_{n,n} = \Var(-s_n T_n) %\\
&= \Var\bigg(\sum_{j=1}^{n-1} s_j T_j\bigg) \\
&= \sum_{j=1}^{n-1} s_j^2 \Var(T_j) + 2 \sum_{1 \le i < j \le n-1} s_i s_j \Cov(T_i,T_j) \\
&= \sum_{j=1}^{n-1} s_j^2 \sigma_{j,j} + 2 \sum_{1 \le i < j \le n-1} s_i s_j \sigma_{i,j}.
\end{align*}
Dividing both sides of this equation by $s_n^2$, we obtain Eq.~(\ref{sigmann}).  
$\qed$

\smallskip

Consider the case in which $s_n$ is small.  Since $s_n^2 < s_n$ then division by $s_n^2$ in Eq.~(\ref{sigmann}) could result in large values of $\sigma_{n,n}$.  Therefore, an insurance plan with a small market share is in danger of having revenue transfer payments which have very large variance.  

This observation also is consistent with empirical data.  The report of the American Academy of Actuaries (2016,~p.~10) shows that the variability of revenue transfers increases substantially as market share decreases.  The report of the Centers for Medicare \& Medicaid Services (2016,~p.~97) indicates that the variability of revenue transfers generally is higher for smaller insurers than for larger insurers.  

\smallskip

We now analyze the nature of changes in $\sigma_{n,n}$ in response to changes in $s_n$.  We are particularly interested in deriving general conditions under which $\sigma_{n,n}$ increases as $s_n$ decreases, because such an event likely will place insurers who are losing enrollees in an even more precarious situation.  

Let $\bLambda$ be the covariance matrix of $(T_1,\ldots,T_{n-1})$; then, $\bLambda$ is a $(n-1) \times (n-1)$ positive semidefinite matrix.  As the $T_i$ are functions of $s_1,\ldots,s_{n-1}$, it will usually be the case that $\bLambda$ also is a function of $s_1,\ldots,s_{n-1}$; nevertheless, we proceed further with the simplifying assumption that $\bLambda$ is not dependent on $s_1,\ldots,s_{n-1}$.  As the ensuing theoretical calculations still lead to conclusions that are consistent with empirical observations, we infer that the dependence of $\bLambda$ on $s_1,\ldots,s_{n-1}$ often is weak.  

Let $\bSigma = (\sigma_{i,j})$ denote the covariance matrix of $(T_1,\ldots,T_n)$ and $\bsigma = (\sigma_{1,n},\ldots,\sigma_{n-1,n})'$ be the $(n-1) \times 1$ column vector containing the covariances between $T_n$ and $(T_1,\ldots,T_{n-1})'$.  Then, 
\begin{equation}
\label{Sigmageneral}
\bSigma = 
\begin{pmatrix}
\bLambda & \bsigma \\
\bsigma' & \sigma_{n,n}
\end{pmatrix}.
\end{equation}
We now establish a necessary and sufficient condition for $\sigma_{n,n}$ to increase as $s_n$ decreases.  

\begin{proposition}
\label{increasingvariance}
For $\sigma_{n,n}$ to increase as $s_n$ decreases, it is necessary and sufficient that $\sigma_{i,n} \le \sigma_{n,n}$ for all $i=1,\ldots,n-1$.  
\end{proposition}

\noindent{\it Proof.}  
Let $\bs = (s_1,\ldots,s_{n-1})'$, written as a $(n-1)\times 1$ column vector.  By Eq.~(\ref{sigmann}), 
\begin{equation}
\label{sigmann2}
\sigma_{n,n} = s_n^{-2} \, \bs' \bLambda \bs.
\end{equation}
Let $\be = (1,\ldots,1)'$  be the $(n-1) \times 1$ column vector whose components all equal $1$.  Then $s_n = 1 - s_1-\cdots-s_{n-1} = 1 - \be'\bs$, and it follows from Eq.~(\ref{sigmann2}) that 
$$
\sigma_{n,n} = (1-\be'\bs)^{-2} \cdot \bs'\bLambda\bs.
$$
Applying the gradient $\bnabla = (\partial/\partial s_1,\ldots,\partial s_{n-1})'$ and simplifying the result, we obtain 
\begin{equation}
\label{gradient_sigmann}
\bnabla \sigma_{n,n} = \frac{2\big[(\bs'\bLambda\bs) \be + (1-\be'\bs) \bLambda\bs\big]}{(1-\be'\bs)^3}.
\end{equation}
By Eq.~(\ref{Sigmageneral}), the $(i,j)$th entry of $\bLambda$ is $\sigma_{i,j}$, $i,j=1,\ldots,n-1$; so, by straightforward matrix multiplication, the $i$th component of the vector $\bLambda\bs$ is $\sum_{j=1}^{n-1} \sigma_{i,j} s_j$.  However, 
\begin{align*}
\sum_{j=1}^{n-1} \sigma_{i,j} s_j &= \sum_{j=1}^{n-1} \Cov(T_i,T_j) s_j 
= \Cov\Big(T_i,\sum_{j=1}^{n-1} s_j T_j\Big).
\end{align*}
Applying (\ref{TnandotherTs}), we obtain 
$$
\sum_{j=1}^{n-1} \sigma_{i,j} s_j = \Cov(T_i,-s_n T_n)  = -s_n \Cov(T_i,T_n) = -s_n \sigma_{i,n},
$$
for all $i=1,\ldots,n-1$; therefore, 
\begin{equation}
\label{Lambdasigma}
\bLambda \bs = - s_n \bsigma.
\end{equation}
We now substitute this result into Eq.~(\ref{gradient_sigmann}), and also substitute $s_n$ for $1 - \be'\bs$ and $s_n^2 \sigma_{n,n}$ for $\bs'\bLambda\bs$.  Then we obtain 
$$
\bnabla \sigma_{n,n} = \frac{2\big(s_n^2 \sigma_{n,n}\be - s_n^2 \bsigma\big)}{s_n^3} 
= \frac{2(\sigma_{n,n} \be -\bsigma)}{s_n}.
$$
Therefore, the $i$th component of $\bnabla \sigma_{n,n}$ is nonnegative if and only if the corresponding component of $\sigma_{n,n} \be -\bsigma$ is nonnegative.  

As the $i$th component of $\sigma_{n,n} \be -\bsigma$ is $\sigma_{n,n} -\sigma_{i,n}$, we find that if $s_n$ decreases, which is equivalent to an increase in at least one of $s_1,\ldots,s_{n-1}$, the condition that $\sigma_{n,n} - \sigma_{i,n} \ge 0$ for all $i$ implies that $\bnabla \sigma_{n,n} \ge 0$, hence $\sigma_{n,n}$ will increase.  
$\qed$

\medskip

Note that it also follows from Eqs.~(\ref{sigmann}), (\ref{Sigmageneral}), and (\ref{Lambdasigma}) that the covariance matrix $\bSigma$ can be expressed entirely in terms of $\bLambda$, $\bs$, and $s_n$ as 
\begin{equation}
\label{SigmaI}
\bSigma = 
\begin{pmatrix}
\bLambda & - s_n^{-1} \bLambda \bs \\
- s_n^{-1} \bs' \bLambda & s_n^{-2} \bs' \bLambda \bs
\end{pmatrix}.
\end{equation}

\smallskip

\begin{example*}{\rm 
Suppose that $n = 3$ and that $s_1 = 0.90$, $s_2 = 0.06$, and $s_3 = 0.04$.  Here, the first insurance plan has a market share of 90\%, which is substantially larger than its competitors.  Although the respective market shares, 6\% and 4\%, of the second and third plans are close to each other, the consequences for the smallest plan are highly negative.  We suppose that 
$$
\bLambda = \left(\begin{array}{rr}
 4.0  & -0.6 \\
-0.6  &  3.0
\end{array}\right),
$$
and then, by Eq.~(\ref{SigmaI}) 
\begin{equation}
\label{Sigma3by3}
\bSigma = \left(\begin{array}{rrr}
 4.0  & -0.6 &  -89.1 \\
-0.6  &  3.0 & 9.0 \\
-89.1 & 9.0 & 1991.25
\end{array}\right).
\end{equation}
Note that $\sigma_{3,3} = 1991.25$, which is substantially larger than $\sigma_{1,1} = 4.0$ and $\sigma_{2,2} = 3.0$.  It is ominous for the smallest plan that the variance of its revenue transfers is enormously greater than the corresponding variances for its larger competitors.  

Also, the conditions of Proposition \ref{increasingvariance} are easily satisfied in this example, so it follows that the variance of revenue transfers for the smallest plan will increase as its market share decreases.  
}\end{example*}

These results indicate that, for any $n$, it is straightforward to find examples for which the variance of the revenue transfers for a given health insurance plan increases as its market share decreases.  We can use the statistical package {\sl R} (R Core Team 2013) to generate examples of $\bSigma$ that satisfy the conditions of Proposition \ref{increasingvariance}, and it is our experience that many such examples exist.

\section{\large Conclusions}
\label{sec:conclusions}

This paper demonstrates three consequences of the revenue transfer methodology in the Affordable Care Act (ACA).  First, we show that there is no practical limit on the magnitude of average revenue transfers.  In the case of insurance marketplace plans with smaller market shares and which typically have less capital, the lack of such a limit increases their risk of insolvency.  

Second, we show that there is no theoretical limit on the variance of revenue transfer payments.  Again in the case of smaller insurance plans, any high volatility in their transfer payments places them at increased risk of insolvency.  

Third, we observe that the variance of the revenue transfer for a given plan often can increase as the plan market share decreases, and we find explicit statistical conditions under which this increase in variance is guaranteed to occur, and we show that the increase holds in cases beyond those conditions.  In the case of smaller plans, the generally higher transfer payments and increased volatility of those payments place them at heightened risk of insolvency.  

%In summary, our results show that larger insurance plans often will have smaller, and less volatile, revenue transfers than smaller insurance plans; and the volatility of an insurance plan's revenue transfers often will increase as its market share decreases.  

In any revision of the ACA, these properties of the revenue transfer methodology need to be addressed.  Further, since our results arise purely as a consequence of the zero sum nature of the revenue transfer process, our results are germane to any health insurance system in which risk equalization is sought by means of zero sum transfer procedures.  

As we noted near the end of Section \ref{sec:thertf}, some ACA participants have proposed that the revenue transfer methodology be revised by placing limits on the magnitudes of transfer amounts as a percentage of premiums.  To explore this approach, suppose that all the insurance plans in a given marketplace were to agree that it is equitable to limit the magnitude of each plan's transfer amount to 50\% of that plan's total premiums.  For simplicity, assume that there are two plans in that marketplace, one having a market share of 20\% and the other with market share 80\%.   Applying the zero-sum condition (\ref{zerosum}), we obtain $0.2\, T_1 + 0.8\, T_2 = 0$; equivalently, $T_1 = -4T_2$.  According to the restrictions on transfer amounts, we also have $|T_1| \, \le 0.5$ and $|T_2| \, \le 0.5$; consequently, $|T_1| \, = 4 \, |T_2| \, \le 0.5$, leading to the restriction $|T_2| \, \le 0.125$.  Therefore, in actuality, the larger plan will have a limit of 12.5\% on the magnitude of its transfer amounts, which is substantially lower than the agreed-to limit of 50\%.  By contrast, the limit on the magnitude of transfers for the smaller plan will remain at 50\%, possibly placing that plan at a disadvantage to the larger plan.  This simple example reveals that the placing of uniform limits on transfer amounts may put smaller plans in a precarious position relative to their larger competitors.

\section*{\large Acknowledgments}
\vspace{-2mm}
We are grateful to John Fricks and Murali Haran for their support during the time that this research was conducted.  

\section*{\large Disclaimer}
\vspace{-2mm}
The authors state that there are no funding sources or other affiliations that may represent a conflict of interest in the results presented in this paper.  The views expressed here are those of the authors and are not necessarily those of their employers.  

\section*{\large Correspondence}
\vspace{-2mm}
Donald Richards, Department of Statistics, 326 Thomas Building, Pennsylvania State University, University Park, PA 16802; E-mail: richards@stat.psu.edu; Tel. (814) 865-3993, Fax. (814) 863-7114.

%\vspace{7mm}

\section*{\large References}

% --- environment for outdenting first line of each paragraph
%     NOTE insert blank line after \begin{reflist}
\newenvironment{reflist}{\begin{list}{}{\itemsep 0mm \parsep 0.2mm
\listparindent -5mm \leftmargin 5mm} \item \ }{\end{list}}

\vspace{-9mm}

\begin{reflist}

\parskip=2.5pt

%\bibitem{ACAinsight}
American Academy of Actuaries 2016.  \href{http://actuary.org/files/imce/Insights_on_the_ACA_Risk_Adjustment_Program.pdf}{\sl Insight on the ACA Risk Adjustment Program}.  Washington, DC.

%\bibitem{Anderson}
%Anderson, T. W. 2003.  {\sl An Introduction to Multivariate Statistical Analysis}, third edition.  Wiley, Hoboken, NJ.

%\bibitem{bryant}
Bryant, M. 2017.  \href{https://www.healthcaredive.com/news/evergreen-health-goes-into-receivership/448609/}{Evergreen Health goes into receivership}.  {\it Healthcare DIVE}, August 4, 2017 (accessed December 28, 2017).

%\bibitem{CHOICES}
Consumers for Health Options, Insurance Coverage In Exchanges in States (CHOICES) 2015.  \href{https://minutemanhealth.org/MinutemanHealth/media/Outreach%20and%20Comms/November,%202015/CHOICES%20White%20Paper%20%e2%80%93%20Technical%20Issues%20with%20Risk%20Adjustment%20and%20Risk%20Corridor%20Programs.pdf}
{Technical Issues with ACA Risk Adjustment and Risk Corridor Programs,} \href{https://minutemanhealth.org/MinutemanHealth/media/Outreach%20and%20Comms/November,%202015/CHOICES%20White%20Paper%20%e2%80%93%20Technical%20Issues%20with%20Risk%20Adjustment%20and%20Risk%20Corridor%20Programs.pdf}
{and Financial Impact on New, Fast-Growing, and Efficient Health Plans} (accessed December 28, 2017).

%\bibitem{cmswhitepaper}
Centers for Medicare \& Medicaid Services, 2016. \href{https://www.cms.gov/CCIIO/Resources/Forms-Reports-and-Other-Resources/Downloads/RA-March-31-White-Paper-032416.pdf}{\sl HHS-Operated Risk Adjustment} \href{https://www.cms.gov/CCIIO/Resources/Forms-Reports-and-Other-Resources/Downloads/RA-March-31-White-Paper-032416.pdf}{\sl Methodology Meeting, Discussion Paper}, March 2016.  Baltimore, MD.

%\bibigtem{Gantz}
Gantz, S. 2016.  \href{http://www.baltimoresun.com/business/bs-bz-evergreen-sues-20160613-story.html}{Evergreen Health Co-op suing federal government over insurance pro-}\href{http://www.baltimoresun.com/business/bs-bz-evergreen-sues-20160613-story.html}{gram}.  {\it The Baltimore Sun}, June 13, 2016 (accessed December 28, 2016). 

%\bibitem{Hermana}
Herman, B. 2016a.  \href{http://www.modernhealthcare.com/article/20160225/NEWS/160229943}{Congress grills CMS over fate of ACA's remaining co-ops}.  {\it Modern Healthcare}, February 25, 2016 (accessed December 28, 2017).

%\bibitem{Hermanb}
Herman, B. 2016b.  \href{http://www.modernhealthcare.com/article/20160630/NEWS/160639997}{ACA's risk adjustment hammers small plans again}. {\it Modern Healthcare}, June 30, 2016 (accessed December 28, 2017).

%\bibitem{Kautteretal}
Kautter, J., Pope, G. C., and Keenan, P. 2014.  \href{https://www.cms.gov/mmrr/Downloads/MMRR2014_004_03_a02.pdf}{Affordable Care Act risk adjustment:} \href{https://www.cms.gov/mmrr/Downloads/MMRR2014_004_03_a02.pdf}{Overview, context, and challenges.}  {\it Medicare \& Medicaid Research Review}, Volume 4, Number 3, 11 pages.

%\bibitem{ACA}
Patient Protection and Affordable Care Act, 2013.  Department of Health and Human Service, Notice of Benefit and Payment Parameters for 2014.  {\sl Federal Register}, Volume 78, Number 47, 15410 (March 11, 2013) (45 CFR Parts 153, 155, 156, et al.).
%{\sl Patient Protection and Affordable Care Act}, Federal Register Volume 78, Number 47 (11 March, 2013), pp.~15430--15432.

%\bibitem{Popeetal}
Pope, G. C., Bachofer, H., Pearlman, A., Kautter, J., Hunter, E., Miller, D., and Keenan, P. 2014.  \href{https://www.cms.gov/mmrr/downloads/mmrr2014_004_03_a04.pdf}{Risk transfer formula for individual and small group markets} \href{https://www.cms.gov/mmrr/downloads/mmrr2014_004_03_a04.pdf}{under the Affordable Care Act}.  {\it Medicare \& Medicaid Research Review}, Volume 4, Number 3, 23 pages.

%\bibitem{RCoreTeam}
R Core Team 2013. {\sl R: A Language and Environment for Statistical Computing}. R Foundation for Statistical Computing, Vienna, Austria. URL: {\color{blue}http://www.R-project.org/}.
  
%\bibitem{Ross2006}
Ross, S. 2006.  {\sl A First Course in Probability}, seventh edition.  Prentice-Hall, NJ.

%\bibitem{Teichert}
Teichert, E. 2016.  \href{http://www.modernhealthcare.com/article/20160706/NEWS/160709977}{HealthyCT crumbles under ACA risk adjustment charges}.  {\it Modern Healthcare}, July 6, 2016 (accessed December 28, 2017).

\end{reflist}

\end{document}